\documentclass[review,12pt]{elsarticle}
\usepackage{amssymb}
\usepackage{amsmath}
\usepackage{amsthm}
\usepackage{lineno}
\usepackage{float}
\usepackage{geometry}
\makeatletter
\def\ps@pprintTitle{%
 \let\@oddhead\@empty
 \let\@evenhead\@empty
 \def\@oddfoot{}%
 \let\@evenfoot\@oddfoot}
\makeatother

\title{Planar 3$\omega$ and 2$\omega$ Method for Increased Sensitivity to Through-Plane Thermal Properties}

\author[]{Darshan Chalise}
\author[]{Brinthan Kanesalingam} 
\author[]{Divya Chalise\corref{cor1}}

\cortext[cor1]{Corresponding author}
\fntext[]{Email: \texttt{divch@stanford.edu}}

\address{Materials Science and Engineering, Stanford University, Stanford, CA 94305, USA}

\journal{International Journal of Heat and Mass Transfer}

\begin{document}

\begin{frontmatter}

\begin{abstract}

Accurately measuring the thermal properties of buried interfaces is crucial for understanding heat transport in multilayered materials, particularly in applications such as batteries and integrated circuits. The conventional 3$\omega$ method, which uses a line heater, has limited sensitivity to through-plane thermal properties due to lateral heat spreading, especially when a highly conductive layer overlays a resistive one. Additionally, when using a line heater, there is a lower limit to the frequency of heating, below which the analytical solution assuming infinite lateral dimension is not valid. This limits analytical interpretation of 2$\omega$ temperature oscillation at lower frequencies where the sensitivity to the conductivity of a buried layer is greater.  To overcome these limitations, we propose a planar 3$\omega$ and 2$\omega$ method that enhances the sensitivity to buried layers by ensuring planar heat flow. We implement this technique using a planar metallic heater across the entire sample and validate it experimentally against the analytical solution using Feldman's algorithm. We also discuss the practical implementation details of the approach, including ensuring uniform current distribution and the required sensor layout. Our results demonstrate improved measurement sensitivity for polymer layers in silicon stacks and buried interfaces in battery electrodes, with sensitivity improvements by factors of 2-5 compared to conventional technique using a line heater. 

\end{abstract}


\end{frontmatter}

\section{Introduction}

The 3$\omega$ \citep{cahill1990thermal} technique using a metallic line heater is a routinely implemented method to measure thermal conductivity of layered structures. The technique involves applying an AC current with a frequency $\omega$  on the metallic line, and the measurement of the resulting temperature oscillations at the frequency 2$\omega$  can be used to determine thermal conductivity or the specific heat capacity of the sample. Applying Feldman’s algorithm \citep{kim1999application} to the metallic line heating allows an analytical prediction of the 2$\omega$ temperature at individual layers in a layered sample. In a conventional 3$\omega$ experiment, the same line heater is used as a sensor. The change in resistance of the line heater due to the temperature oscillations at 2$\omega$ results in a change in the voltage dropped across the sample at 3$\omega$, which can be detected using a lock-in amplifier- thus resulting in the name,  the 3$\omega$ technique. Alternatively, the sensor can be placed in a different layer, and when a DC current is applied on the sensor, the 2$\omega$ temperature oscillations result in a voltage drop across the sensor at frequency 2$\omega$ \citep{chalise2024depth}.

The 2$\omega$ temperature response in a conventional 3$\omega$ experiment with a line heater is, in general, sensitivity to both the through-plane ($\kappa_z$) and the in-plane ($\kappa_x$ or $\kappa_y$) thermal conductivity of individual layers. By changing the width of the metallic heater/sensor, the sensitivities to the through-plane and in-plane conductivities can be altered \citep{mishra20153}.  However, theoretically, the sensitivity to in-plane conductivity remains non-zero as long as the heater width is smaller than the lateral dimensions of the sample. On the other hand, the analytical solution to the 2$\omega$ temperature response \citep{kim1999application}, which allows for simple fitting of the data to measure the thermal properties of individual layers, assumes that the lateral dimension of the sample is infinite. If the lateral dimension of the sample is not significantly greater than the thermal penetration depth of heating in the lateral direction ($\sqrt{\frac{D_{x\text{ or } y}}{2\omega}}$), the analytical solution to 3$\omega$ with line heater/sensor cannot be applied. Here, $D_{x\text{ or } y}$ represents the thermal diffusion coefficient along the direction of interest, x or y, and $\omega$ is the frequency at which the AC current is applied. This sets a lower limit on the frequency, below which the analytical solution to 2$\omega$ cannot be applied when using a line heater. This is given by 

\begin{equation}
     \frac{l_{x\text{ or } y}}{2}-b = 2 \sqrt{\frac{D_{x\text{ or } y}}{2 \omega_{min}}}.
     \label{equation1}
\end{equation}

Here, $l_{x\text{ or } y}$ is the lateral width of the sample in a particular direction and $b$ is the half-width of the heater.

We use the prefactor 2 in the right-hand side of eq.~\ref{equation1} assuming, if the lateral width of the sample in either direction is at least 2 times greater than the thermal penetration depth, the analytical solution assuming an infinite lateral dimension is not significantly incorrect. We also assume that the heater is placed in the middle of the sample, thus using $\frac{l_{x\text{ or } y}}{2}$ to describe the lateral width of the sample. Along the direction of the length of the heater, eq.~\ref{equation1} should be modified to 

\begin{equation}
    \frac{l_{x\text{ or } y}}2-a = 2 \sqrt{\frac{D_{x\text{ or } y}}{2\omega_{min}}},
    \label{equation2}
\end{equation}

where \textit{a} is the length of the heater.  

For samples involving stacking of a high lateral thermal conductivity layer over a low thermal conductivity layer (resistive layer), the sensitivity of conventional 3$\omega$ to the thermal conductivity of the resistive layer is greatly limited above the minimum acceptable frequency (see for example the sensitivity of conventional 3$\omega$ in the geometry in section~\ref{3DIC}). However, accurate measurement of thermal conductivity of such deeply buried layer has technological relevance in important systems, the examples for which include the measurement of  interface resistance of a battery \citep{chalise2023using,lubner2020identification} with a copper current collector as a top layer or the measurement of the thermal conductivity of a polymer layer in a high-bandwidth memory stack \citep{jun2017hbm} with silicon as the top layer.

Application of a 3$\omega$ or a 2$\omega$ heating/detection with a wide heater and at a frequency lower than $\omega_{min}$ can be implemented to enhance the sensitivity of the measurement to the conductivity of the buried high resistance layer. However, analytically solving for the 2$\omega$ temperature response based on the thermal transport coefficients of individual layers is then significantly challenging. Finite element methods can be implemented to calculate the temperature response and obtain a best fit to thermal properties that results in a temperature response recorded in the experiment. However, setting up finite element simulation for every individual sample, heating and detection geometries and solving for the thermal properties resulting in the best fit is both computationally demanding and not general.

In the specific case where the heater is applied to the entirety of the sample and the sample is of the same lateral dimension in all the layers, the heat flow will be planar along the depth of the sample. In this case, the analytical solution to relating the  2$\omega$ temperature oscillations to the thermal transport properties of individual layers is possible. Indeed, such solution has been provided by Feldman \citep{feldman1999algorithm} and is only dependent on the transport properties in the depth direction. Implementing a 3$\omega$ scheme (or a 2$\omega$ scheme if the heater and a sensor are not the same layers) with this planar heat flow  greatly enhances the sensitivity of the measurement to thermal conductivity of the buried layers in the case a highly conductive layer is above a resistive layer. In addition, as the restriction placed on minimum frequency by lateral dimension by eq.~\ref{equation1} is no longer applicable, measurements at much smaller frequenciesm allowing significantly longer thermal penetration depth along the stacking direction, can still be interpreted analytically.

In this paper, we present the method to implement planar 3$\omega$ and 2$\omega$ schemes for planar heating geometry. We begin with a brief review of the solution to the 2$\omega$ temperature oscillations using the Feldman algorithm in a planar heating geometry generalized to an arbitrary layer for heating and sensing. We then present a practical method for implementing the scheme and discuss the caveats when carrying out the implementation. Finally, we demonstrate how this planar 3$\omega$ measurement allows analytical determination of the thermal conductivity of buried polymer layers in a silicon stack which would not be possible with a conventional 3$\omega$ measurement. 

\section{Generalized Feldman's solution for planar heating} 
\label{feldman}

Feldman \citep{feldman1999algorithm} presents the temperature solution for a planar sinusoidal heat source at an arbitrary location in a stack. 

The temperature at the two boundaries of the stack T$_0$ and T$_{N+1}$ can be calculated as 

\begin{eqnarray}
    T_{0} &= \frac{q}{2\gamma_j} \frac{B^{+}+B^{-}}{A^{+}B^{-} - A^{-}B^{+}}\\
    T_{N+1} &= \frac{q}{2\gamma_j} \frac{A^{+}+A^{-}}{A^{+}B^{-} - A^{-}B^{+}}
\end{eqnarray}

Where $q$ is the magnitude of the heat generation per square unit area, $\gamma_j$ is given by  $\gamma_j = u_j k_j$, where   $k_j$ is the thermal conductivity of the $j^{th}$ layer where the heat is generated and $u_j = -i \frac{\omega}{D_j}$.  Here, $\omega$ is the frequency of the heat generation and $D_j$ is the thermal diffusivity of the $j^{th}$ layer. 

A and B are vectors related to thermal transport across the boundary and within mediums of the stack. 

\begin{equation}
\begin{aligned}
\mathbf{A} &= U_j(\xi) \times \Gamma_{j,j-1} \times \dots \times \Gamma_{3,2} \times U_{2}(L_2) \times \Gamma_{2,1} \times U_1(L_1) \times \Gamma_{1,0} \times 
\begin{pmatrix} 
1 \\ 
0 
\end{pmatrix};
\end{aligned}
\end{equation}

\begin{equation}
\begin{aligned}
\mathbf{B} &= U_j(\xi - L_j) \times \Gamma_{j,j+1} \times U_{j+1}(-L_{j+1}) \times \Gamma_{j+1,j+2} \times \dots \\
&\quad \dots \times U_{N-1}(-L_{N-1}) \times \Gamma_{N-1,N} \times U_N(-L_N) \times \Gamma_{N,N+1} \times 
\begin{pmatrix} 
0 \\ 
1 
\end{pmatrix};
\end{aligned}
\end{equation}

Where $\xi$ is the location of the heating source in $j^{th}$ layer of the stack, $ \mathbf{U}_a(x) =
\begin{pmatrix} 
e^{u_a x} & 0 \\ 
0 & e^{-u_a x} 
\end{pmatrix}
$, $
\Gamma_{ba} = \frac{1}{2\gamma_b} 
\begin{pmatrix} 
\gamma_b + \gamma_a & \gamma_b - \gamma_a \\ 
\gamma_b - \gamma_a & \gamma_b + \gamma_a 
\end{pmatrix}
$ and $L_j$ is the thickness of the $j^{th}$ layer. 
\vspace{1em}

If the planar heater has a finite thickness, the temperature solution presented by Feldman can be used as a Green’s function solution and integrated over the thickness of the heater to calculate the temperature T$_0$ and T$_{N+1}$ at the boundaries \citep{mishra20153}. 

The Green’s function for the temperature sensed at any location within the stack can be calculated from one of the boundary temperatures using vector relationships for thermal transport within a medium (Equation 6 in Ref.~\citep{feldman1999algorithm}) and across a boundary (Equation 8 in Ref.~\citep{feldman1999algorithm}), provided that there is no heat source between the boundary layer and the sensor layer. Practically, if the heat source lies at the left (Figure 2 of Ref.~\citep{feldman1999algorithm}) of the sensor, then the temperature at the sensor can be calculated by applying the vector relationship on T$_{N+1}$. Similarly, if the heat source is at the right side of the sensor, then it can be calculated from T$_0$. The temperature (or Green’s function) thus calculated is a two-dimensional vector with components (T$_{j+}$ and T$_{j-}$), which when evaluated at the left boundary ($z=0$) of the sensor layer gives the measured complex temperature as T=T$_{j+}$+T$_{j-}$. 

If the heater and the sensor layer are same, assuming that there is no thermal transport lag within the sensor (a reasonable assumption for a metallic heater/sensor), the sensor's measured temperature can be determined by modeling the sensor layer as a thin, non-heat-generating layer positioned at either the leftmost or rightmost boundary of the heater and with the same thermal properties.

\section{Implementation of 3$\omega$ and 2$\omega$ scheme for planar heating and detection}
\label{impl}

\begin{figure}[H]
\centering
\includegraphics[width=\textwidth]{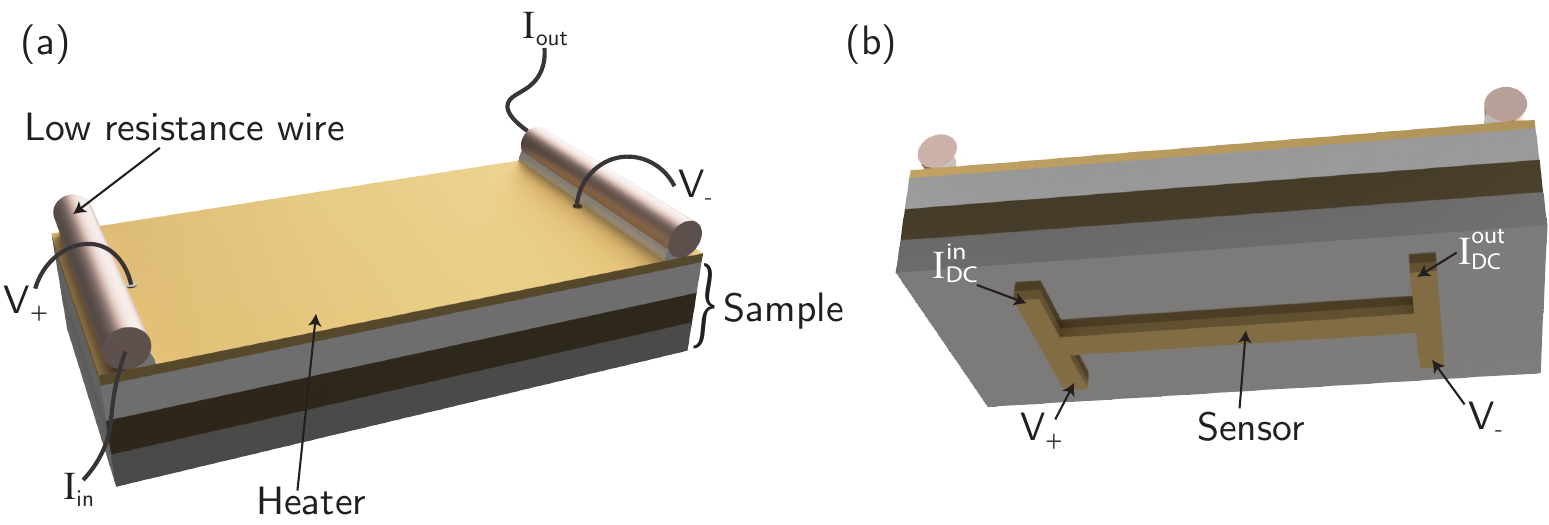}
\caption{Implementation of planar 3$\omega$ and 2$\omega$ methods in a sample. (a) shows the implementation of 3$\omega$, where a thick, low resistance wire attached to the planar heater via conductive silver epoxy for spreading the input 1$\omega$ current is used. The leads to read out the 3$\omega$ voltage across the planar heater are attached slightly away from the current spreading wire. (b) shows the additional deposition of the line sensor needed for the detection of the 2$\omega$ voltage from a different side. The sensor includes leads for DC current flow and for 2$\omega$ voltage detection. }
\label{fig1}
\end{figure}

The schematic describing the implementation of a planar 3$\omega$ heating/detection is shown in Fig.~\ref{fig1}(a). The same implementation in the 2$\omega$ scheme with the heater on a different side than the sample is shown in  Fig.~\ref{fig1}(b). We primarily describe the implementation for a sample whose geometry can be approximated by a rectangular shape.

The implementation of planar 3$\omega$ and 2$\omega$ requires ensuring a planar heating geometry. This is accomplished by first depositing a planar metallic heater across the entire sample. As the 2$\omega$ temperature rises and consequently the 2$\omega$  or 3$\omega$ voltage response scales linearly with the resistance of the heater, it is important to ensure that the resistance across the heater is large enough for a large signal. We have observed that for a $\sim$ 1~cm x 1~cm sample, a silver, gold, or platinum heater with a thickness of $\sim$ 50~nm provides a high enough resistance for a measurement.

Once a planar heater is deposited, it is imperative to ensure that the current flowing across the entire cross-section of the sample is uniform. We observed that connecting the current leads at any two ends of the sample, typical for a conventional 3$\omega$ implementation, results in an incorrect determination of the thermal parameters as the current across the heater is not uniform. 

A uniform current across the heater is ensured using a low-resistance wire attached to the entirety of the two ends of the heater, as shown in Fig~\ref{fig1}. In our experiment, we attach thick copper wires across the heater and use silver epoxy to ensure proper electrical contact between the copper wire and the heater. The electrical resistance of the copper and silver epoxy should be significantly smaller than the resistance of the heater to ensure that the copper wire/silver epoxy spread the current across the heater uniformly.

Once the spreading of the current across the heater is ensured, the 3$\omega$ detection can be carried out by placing the leads for the voltage sensor anywhere in the sample in between the current spreading wire/epoxy. One should ensure, when calculating the temperature response, only the area of the heater in between the voltage leads is taken into account.

For the 2$\omega$ implementation, we have observed that placing a sensor with a smaller form factor is more efficient. Unlike in the 3$\omega$ scheme, where only the heating in between the voltage leads need to be considered and the heating per unit area has less ambiguity, in the 2$\omega$ scheme, there could be ambiguity in determining the heating per unit area due to the spread of the epoxy placed for contact. In order to avoid errors due to this ambiguity, the epoxy should be applied as minimally as possible while ensuring uniformity and minimizing the resistance from the epoxy.

To increase the resistance of the wire, and therefore increase the 2$\omega$ temperature response, as well as to allow for planar heating of the sample, we also tried using a serpentine heater instead of a planar heater. The serpentine heater had a  with spacing smaller thickness of individual wire. We observed that the determination of the thermal parameters using the 3$\omega$ scheme when using a serpentine heater is erroneous, primarily the out-of-phase response. We hypothesize that this is because the density of serpentine heating, which are not completely planar, results in this error. For a 2$\omega$ scheme, the reconstruction when using a serpentine heater is less erroneous perhaps because the sensor samples the average temperature response. We point the readers to the discussion in \ref{app1} for errors observed when using a serpentine heater. However, in general, we suggest applying the planar heating scheme described in Fig~\ref{fig1} in general as this avoids the issues encountered with a serpentine heater.

\section{Experimental Details}
\label{exp}

All the experiments were performed by depositing 50-70 nm platinum or silver heaters and sensors on the sample with $\sim$10 nm of Cr or Ti as the adhesion layer.

The circuit design for 3$\omega$ has been detailed in Ref.~\citep{cahill1990thermal}. We use a SR 830 lock-in amplifier as both a current source and for the detection of the 2$\omega$ or 3$\omega$ voltage drop across the sensor. We use the reference channel of the SR830 lock-in amplifier for the input current in the heater. Once the 1$\omega$ across the heater is balanced using a variable resistor, we note down the 1$\omega$ voltage across the heater to calculate the 1$\omega$ current across the heater, using the steady-state resistance of the heater. Finally, the 3$\omega$ signal is measured using the 3rd harmonic of the difference between the voltage dropped across the heater/sensor and the variable resistor.

For 2$\omega$ measurements, we use the DC output of the lock-in amplifier and follow a similar cancellation scheme for a 3$\omega$ detection and perform the measurement using the $2^{nd}$ harmonic of the difference between the voltage dropped across the sensor and the variable resistor.

The thermal conductivity and the heat capacity of the Kapton film used for the verification experiments have been reported in Ref.~\citep{chalise2024depth}.

For all the measurements, we use a Keysight 34401A digital multimeter and a temperature-controlled hot plate to calibrate the temperature coefficient of resistance (TCR) of heaters and sensors. 

\section{Verification of the plane 3$\omega$ and 2$\omega$ methods}
\label{verification}

\begin{figure}[H]
\centering
\includegraphics[width=\textwidth]{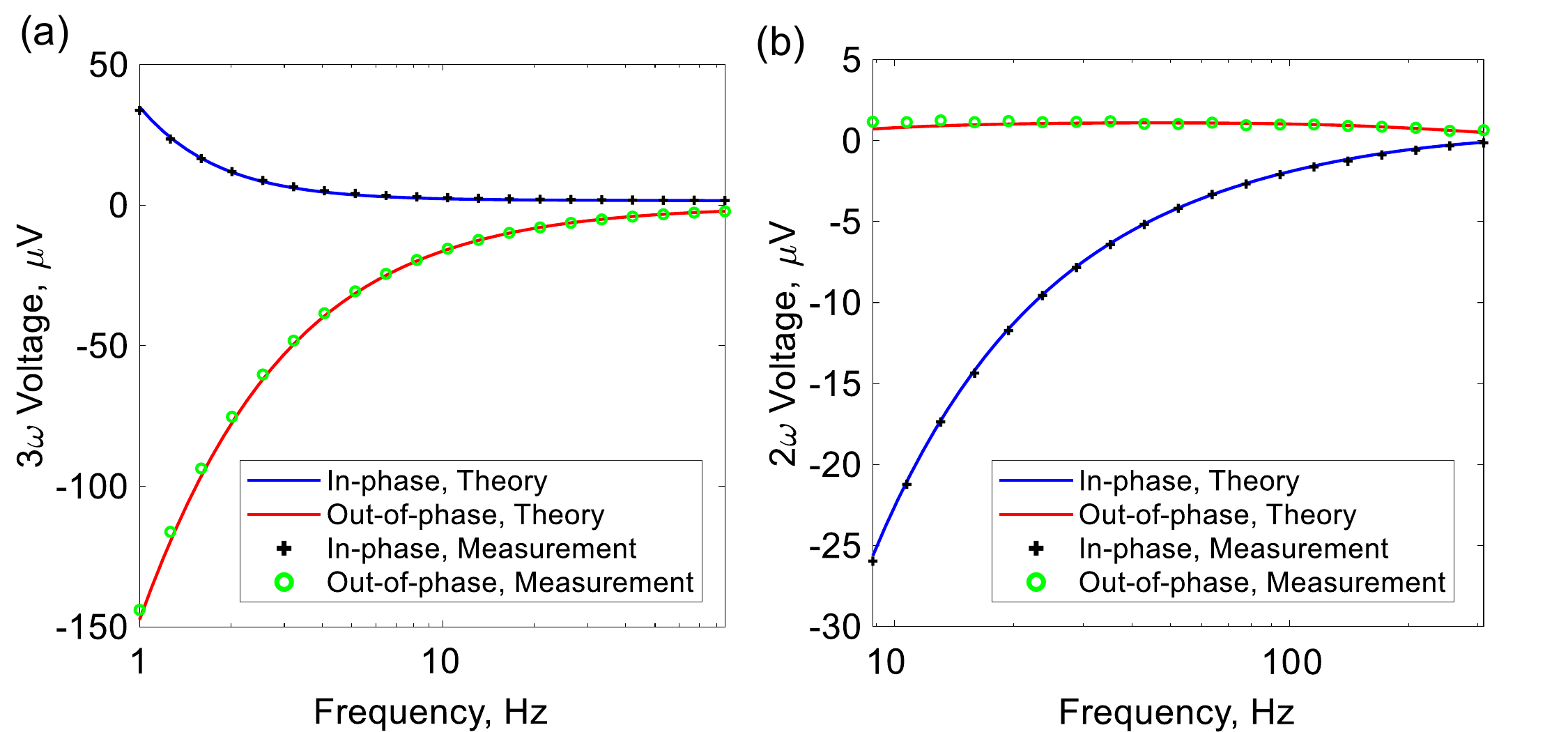}
\caption{Planar thermal measurements on Kapton film. (a) 3$\omega$ method with heater as the sensor. (b) 2$\omega$ method with separate heater and sensor. Solid lines show theoretical predictions using Feldman's method, matching experimental in-phase and out-of-phase voltages. In 3$\omega$ (a), the in-phase voltage is primarily related to the thermal conductivity, while the out-of-phase is related to the heat capacity. For 2$\omega$ (b), the relationship is reversed due to the DC current in the sensor, with in-phase voltage sensitive to the heat capacity and the out-of-phase to the thermal conductivity.}
\label{fig2}
\end{figure}

Figure~\ref{fig2} (a) and \ref{fig2} (b) show the planar 3$\omega$ and planar 2$\omega$ measurements on kapton film whose thermal properties have been previously determined in Ref.~\citep{chalise2024depth} with the heater acting as the sensor in the planar 3$\omega$ method and the heater and the sensor placed in two sides of the film in the planar 2$\omega$ method. As seen, the theoretical predictions from the Feldman’s method (solid lines) with the pre-determined thermal properties match perfectly with the experimentally measured in-phase and out-of-phase voltages for both the 3$\omega$ and 2$\omega$ measurements. For the 2$\omega$ measurements, as the sensor has DC current across it, the phase of the 2$\omega$ voltage matches with the phase of the 2$\omega$ temperature (and resistance) oscillations. Therefore, the in-phase voltage relates to the primary component of the temperature rise, which is sensitive to the heating magnitude and the specific heat capacity of the material. The out-of-phase voltage is related to the thermal lag between the heat source and the temperature rise, and is therefore primarily sensitive to the thermal conductivity. For the 3$\omega$ measurement, the sinusoidal (1$\omega$) current is used to sense the 2$\omega$ temperature oscillations. Thus, it induces a 90$^\circ$ phase flip between the sensed 3$\omega$ voltage and the 2$\omega$ temperature oscillation. Therefore, the in-phase voltage is primarily sensitive to the thermal conductivity while the out-of-phase voltage is primarily sensitive to the specific heat capacity of the material. 

\section{Implementation of the planar 3$\omega$ method in a silicon-polymer stack}
\label{3DIC}
\begin{figure}[H]
\centering
\includegraphics[width=\textwidth]{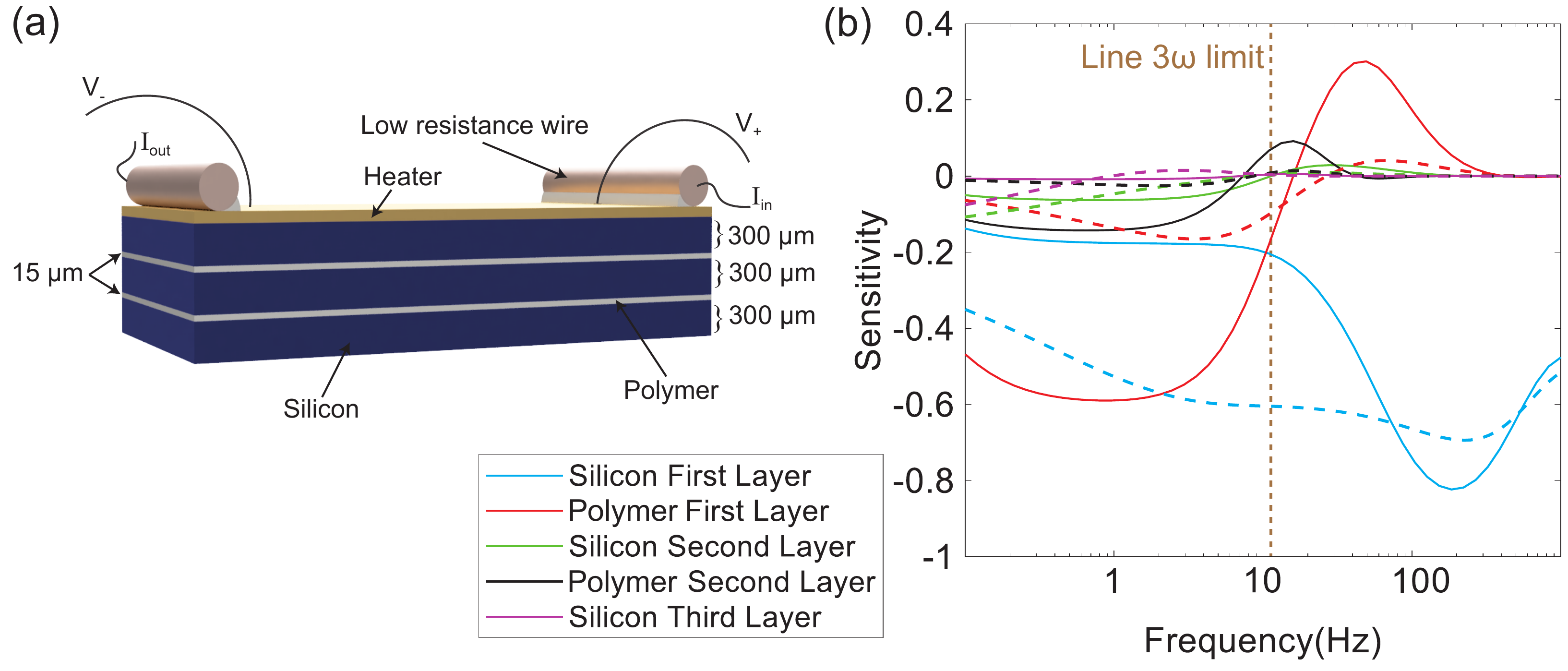}
\caption{Measurement sensitivity as a function of frequency for thermal conductivity and thermal interface resistance using the traditional (line) 3$\omega$ method (dashed lines) and planar 3$\omega$ method (solid lines) for silicon and buried polymer layers in a stack. (a) Arrangement of the stack used in our analysis. The stack consists of three silicon layers, each with thickness of 300 $\mu$m and two polymer layers in between the silicon layers, each with thickness 15 $\mu$m. The heater/sensor used for 3$\omega$ is deposited on the top silicon. The figure shows the implementation of planar 3$\omega$ with the current spreading wire at the top and a heater deposited throughout the sample. For traditional line 3$\omega$, the deposited heater is instead a line heater with 4 leads for passing the current and reading out the voltage. The lateral width of the sample used in our analysis is 7 mm (b) Plot showing the sensitivities to the thermal conductivities of each  layer using the line 3$\omega$ method (dashed lines) and the planar 3$\omega$ method (solid lines). The vertical line at 13.7 Hz shows the low frequency limit of the analytical solution to line 3$\omega$ corresponding to the lateral width of the sample of 7 mm. Legend is shown on the left side of the plot.}
\label{si}
\end{figure}

To demonstrate the advantage of using the planar 3$\omega$ scheme when layers with high thermal resistance are buried below layers with high thermal conductivity, we present the comparison of the measurement of the thermal conductivity of buried polymer layers in a stack as shown in Figure~\ref{si} (a). The stack consists of 3 layers of undoped silicon (with thermal conductivity of 140 W/mK and volumetric specific heat capacity of 1.65 MJ/m$^{3}$K), each with a thickness of 300 $\mu$m. In between the silicon layers is a layer of polymer that is 15 $\mu$m thick. This is representative of the layering in a vertically stacked integrated circuit \citep{jun2017hbm}. We use the thermal conductivity of the polymer as 1.3 W/mK (the thermal conductivity of a thermally conductive epoxy) and a specific heat capacity of 1.6 MJ/m$^{3}$K for our computation. 

Figure~\ref{si} (b) presents the comparison of the frequency dependent sensitivity of the measurement to thermal conductivity of all the layers when using 3$\omega$ with a 150 $\mu$m line heater (dashed lines) and planar heater (solid lines). As seen in the figure, the sensitivity in the measurement using a line heater is primarily dominated by the thermal conductivity of the top silicon layer at all frequencies. This is because the heat has a tendency to spread across the top, conductive silicon layer. Moreover, if there is a finite size of the sample in the lateral dimension, this places a minimum frequency of the measurement below which the analytical solution cannot be applied, given by equation~\ref{equation1}. In the example presented in Figure~\ref{si}, we use the lateral width of the sample as 10 mm (typical of a 3D integrated circuit), which places a minimum frequency of 13.7 Hz. This means, when using the line heater, the maximum sensitivity one can obtain for the first buried polymer layer is ~ 0.1, and there is practically no sensitivity for the second polymer layer.

In contrast, when using the 3$\omega$ method with a planar heater deposited across the sample, the sensitivity is primarily dominated by the highly resistive first polymer layer below 10 Hz. Additionally, since the analytical solution can be applied at all frequencies, at frequencies below 1 Hz, the sensitivity to the thermal conductivity of the 2nd polymer layer is also greater than 0.1. This means, it is also possible to determine the conductivity  of the 2nd polymer layer by applying the analytical solution using the planar heating approach.

\section{Implementation of the planar 3$\omega$ method in a battery stack}
\label{BS}

\begin{figure}[H]
\centering
\includegraphics[width=\textwidth]{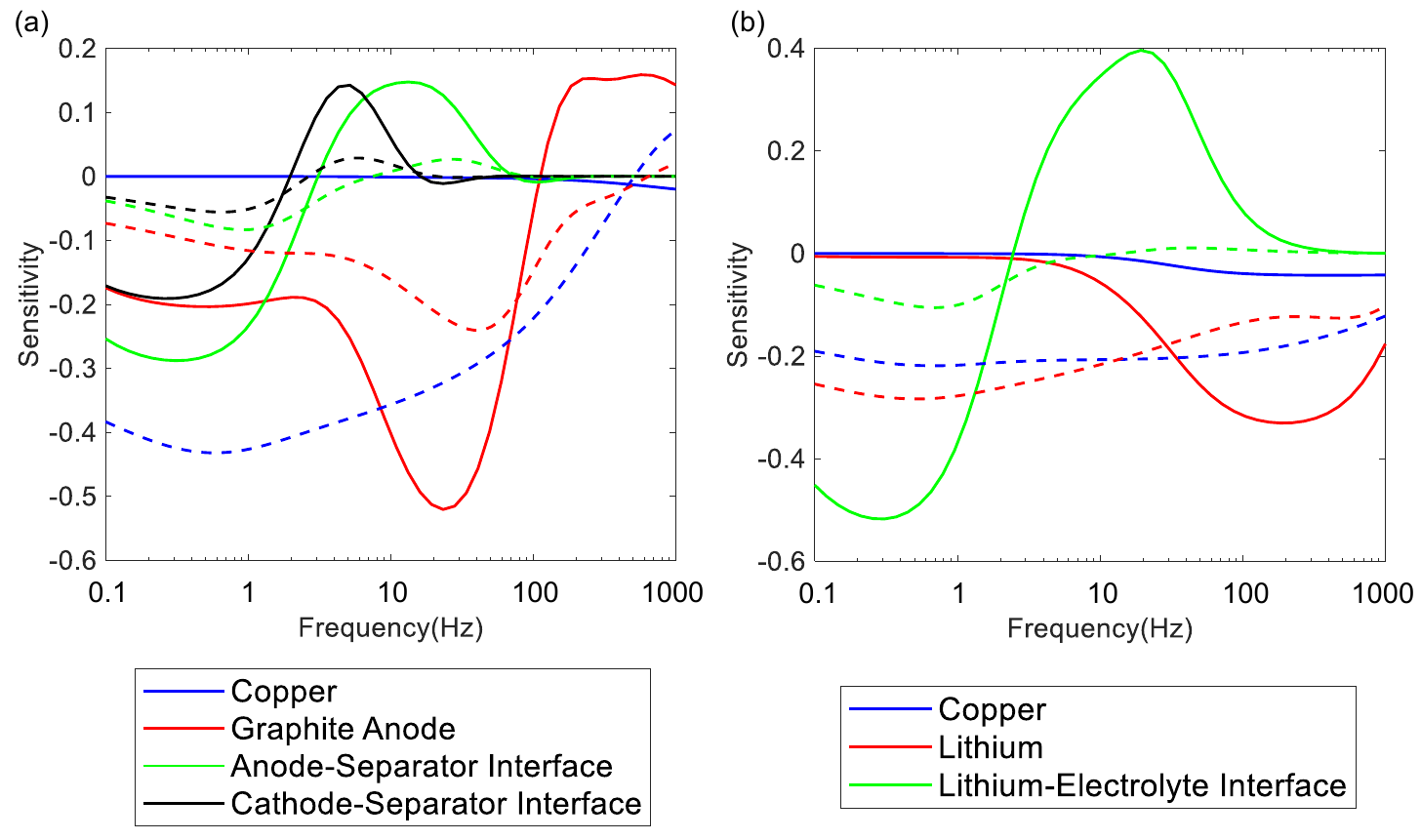}

\caption{Measurement sensitivity as a function of frequency for thermal conductivity and thermal interface resistance using the traditional (line) 3$\omega$ method (dashed lines) and planar 3$\omega$ method (solid lines) for components and interfaces in a lithium-ion battery (a) and a symmetric solid-state battery (b). For the lithium-ion battery (a), the traditional 3$\omega$ method is primarily sensitive to the copper film's thermal conductivity, with limited sensitivity to the graphite electrode and the anode-separator and cathode-separator interfaces. In contrast, the planar 3$\omega$ method shows increased sensitivity to the anode and anode-separator and cathode-separator interfaces by factors of 2, 3.5, and 3.8 respectively, while being less sensitive to the copper film's conductivity. For a symmetric lithium metal solid-state cell (b), the traditional 3$\omega$  method is most sensitive to the copper current collector and lithium foil anode, with lithium metal-electrolyte interface sensitivity below 0.1. The planar method improves interface sensitivity by a factor of 5.}
\label{fig3}
\end{figure}

An example of the advantage of the planar 3$\omega$ method is evident in the depth-resolved thermal metrology of batteries. The thermal interface resistance of the electrode-electrolyte interface \citep{chalise2023using,zeng2023nonintrusive} as well as the thermal conductivity of the electrode layers \citep{zeng2021operando} have been related to electrochemical phenomena such as electrochemical degradation and lithiation in lithium-ion and solid-state batteries. This series of works has established thermal metrology, particularly the 3$\omega$ method, as a possible diagnostic tool for electrochemical changes within batteries. However, due to the presence of metallic current collectors (copper and aluminum) between the 3$\omega$ sensor and the electrodes/interfaces of interest, the heat from the 3$\omega$ sensor preferentially flows in the in-plane direction of the metallic current collectors, affecting the measurement sensitivity for the interface and electrode thermal properties. With the planar 3$\omega$ method, however, the heat flow is restricted to the through-plane direction, maximizing the measurement sensitivity to the interfaces and electrodes. Figure~\ref{fig3} shows the in-phase measurement sensitivity to the thermal conductivity (or the thermal interface resistance) for the copper current collector, graphite anode, the anode-separator interface, and the cathode-separator interface respectively for a typical lithium-ion battery as a function of the measurement frequency. The measurement sensitivities from the traditional (line) 3$\omega$ method are shown in dashed lines while those from the planar 3$\omega$ method are shown in solid lines. As seen, the line 3$\omega$ method is mostly sensitive to the thermal conductivity of the copper film, while the measurement sensitivities to the graphite electrode, as well as the interfaces, are significantly compromised. The measurement sensitivities for anode-separator and cathode-separator interfaces are less than 0.1 and therefore not within the measurement resolution. On the contrary, the planar 3$\omega$ method is not sensitive to the conductivity of the copper film and is more sensitive to the thermal properties of the anode and interfaces of interest by a factor of 2, 3.5, and 3.8 respectively. Similarly, for a symmetric lithium metal solid-state cell with a lithium foil as the anode material and a ceramic solid-state electrolyte, the traditional 3$\omega$ method is mostly sensitive to the thermal conductivity of the copper current collector and the lithium foil anode, as shown by the dashed lines in Figure~\ref{fig3} (b). The measurement sensitivity to the solid-solid interface between the lithium metal anode and the solid-state electrolyte is less than 0.1. However, with the implementation of the planar 3$\omega$ method, the measurement sensitivity of the interface increases to 0.5, which is an improvement by a factor of 5.

\section{Discussion and Conclusions}
\label{Discussion}

In this work, we have presented the analytical solution and the practical approach to implement the planar 3$\omega$ and 2$\omega$ methods, which enable measuring the thermal properties of layered structures with increased sensitivity to the through-plane properties. Compared to the conventional line heater technique, the planar approach significantly improves sensitivity to the thermal properties of buried layers and interfaces in multilayer stacks with a top conductive layer. We expect this approach will improve thermal metrology in vertically stacked integrated circuits where both thermometry and thermal conductivity measurement of the polymer layer are traditionally challenging \citep{chalise2022temperature}\citep{chalise2023electron}. In batteries, where the thermal conductivity of the electrode is an indicator of the lithiation and the thermal resistance of the electrolye-electrode interface is an indicator of degradation, we expect the method to significantly improve the diagnosis of the lithiation state as well as the degradation. 

Although we do not explicitly calculate the sensitivities offered by the 2$\omega$ implementation, the sensitivity to an individual layer can be improved by using a 2$\omega$ sensor next to the layer of interest instead of using the 3$\omega$ detection from the top.  Therefore, we also expect the 2$\omega$ implementation discussed in Section 3 to be useful to the readers.  

One limitation of the implementations discussed in this work is that it requires every individual layer to be homogeneous across the lateral dimension. If the sample is significantly inhomogeneous across the lateral dimension, the analytical implementation discussed here cannot be directly used.  In those cases, a computation solution has to be applied to understand the 2$\omega$ temperature oscillation across the sensor.

\section*{Acknowledgments}

The authors acknowledge Prof. David Cahill (UIUC) and Prof. Arun Majumdar (Stanford) for their valuable discussions related to this work and Thomas E. Carver (Stanford) for assistance with sensor fabrication. The fabrication of the heater/sensor was supported by Prof. Majumdar’s gift fund. Part of this work was performed at the Stanford Nano Shared Facilities (SNSF), supported by the National Science Foundation under award ECCS-2026822.

\bibliographystyle{ieeetr} 
\bibliography{bibliography}

\appendix
\setcounter{figure}{0}

\section{Implementation of the planar 3$\omega$ method with a serpentine heater}
\label{app1}

\begin{figure}[H]
\centering
\includegraphics[width=0.7\textwidth]{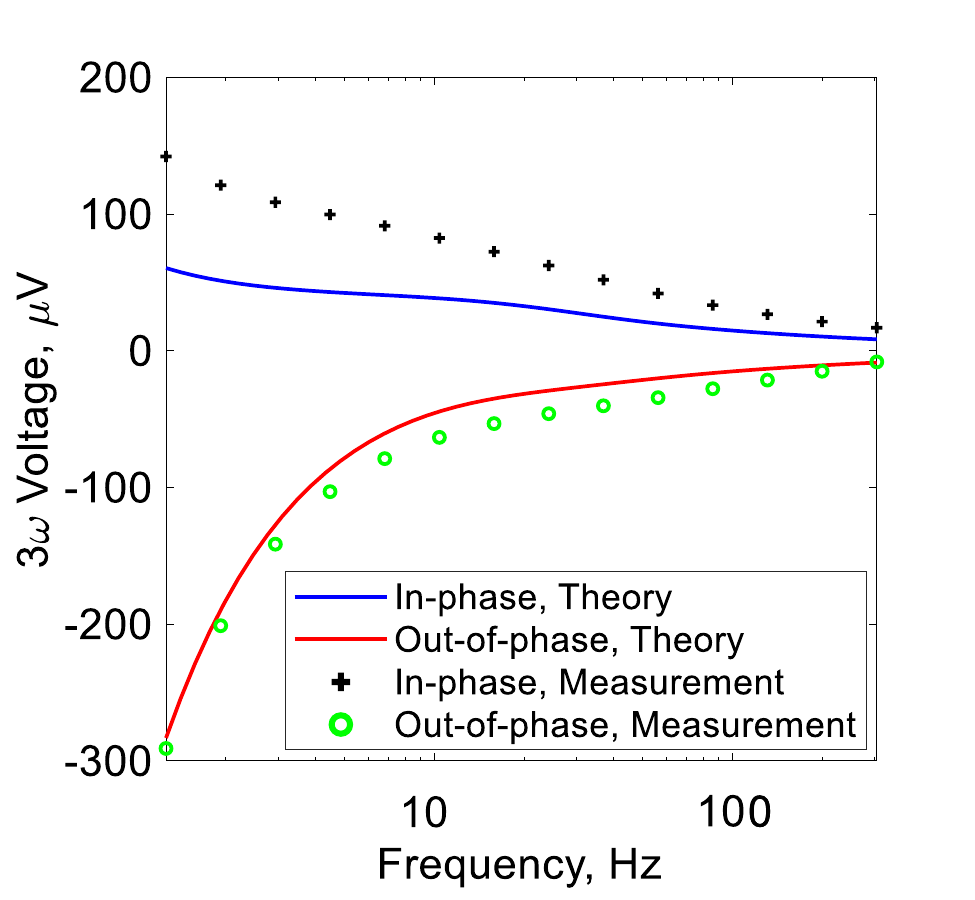}
\caption{Implementation of the planar 3$\omega$ method with a serpentine heater. The experimentally measured out-of-phase term related to the specific heat matches reasonably well with the prediction from the Feldman’s method while the in-phase term does not.}
\label{a1}
\end{figure}

Instead of using a planar heater with current flowing uniformy in the plane, planar heating can also be tried with a patterned serpentine heater. However, with this implementation in the planar 3$\omega$ method, with the heater acting as the sensor, we have observed that in phase and out of phase temperatures predicted from Feldman’s solution to planar heating do not match the experimentally measured temperatures. In particular, the in-phase temperature deviates significantly from the theoretical prediction. This discrepancy is shown in Figure~\ref{a1} where a serpentine heater/sensor was used to measure the frequency domain temperature rise on the same kapton film used to verify the planar 3$\omega$ and 2$\omega$ methods in Figure~\ref{a1}. We believe this deviation comes from localized heating in the serpentine heater/sensor which makes the temperature at the heater/sensor higher than the average temperature of the plane at which the heater/sensor is located. For the 2$\omega$ method however, with a DC line sensor running across the sample, it is previously observed \citep{chalise2024depth} that the 2$\omega$ temperature measured matches well with the prediction from the Feldman’s method, most likely due to the averaging out of the temperature at the sensor which runs across the projected area of the serpentine heater.

\end{document}